\documentclass[a4paper,nofootinbib,useAMS,usenatbib]{mn2e}
\usepackage{amssymb}
\usepackage{amsmath}
\usepackage{graphicx,longtable,mathrsfs,color,array}
\usepackage{epsfig,subfigure,placeins,float} 
\usepackage{mathrsfs}
\usepackage{longtable}
\usepackage[usenames,dvipsnames]{xcolor}
\usepackage{mathtools}
\usepackage{relsize}
\usepackage{geometry}
\usepackage{graphicx}
 \usepackage{hyperref}

\providecommand{\eprint}[1]{\href{http://arxiv.org/abs/#1}{#1}}
\providecommand{\adsurl}[1]{\href{#1}{ADS}}

\usepackage{color}

\newcommand{\apj}{ApJ}			
\newcommand{\apjs}{ApJS}		
\newcommand{\jcap}{JCAP}		
\newcommand{\nar}{New Astronomy Reviews}
\newcommand{\prd}{Phys.~Rev.~D}		
\newcommand{\prl}{Phys.~Rev.~Lett.}	
\newcommand{\pasp}{PASP}		
\newcommand{\nat}{Nature}		

\newcommand*{\rom}[1]{\expandafter\@slowromancap\romannumeral #1@}
\makeatother

\title[Axion dark matter, solitons, and the cusp-core problem]{Axion dark matter, solitons, and the cusp-core problem}
\author[D. J. E. Marsh and A. R. Pop]{David J. E. Marsh$^{1}$\thanks{E-mail: dmarsh@perimeterinstitute.ca} and Ana-Roxana Pop$^{2}$\thanks{E-mail: pop@princeton.edu}\\
$^{1}$Perimeter Institute, 31 Caroline St N, Waterloo, ON, N2L 6B9, Canada \\
$^{2}$Department of Physics, Princeton University, Princeton, NJ 08544, USA}
\begin{document}

\date{Draft version: \today}

\pagerange{\pageref{firstpage}--\pageref{lastpage}} \pubyear{2014}

\maketitle

\label{firstpage}

\begin{abstract}
Self-gravitating bosonic fields can support stable and localised (solitonic) field configurations. Such solitons should be ubiquitous in models of axion dark matter, with their characteristic mass and size depending on some inverse power of the axion mass, $m_a$. Using a scaling symmetry and the uncertainty principle, the soliton core size can be related to the central density and axion mass in a universal way. Solitons have a constant central density due to pressure-support, unlike the cuspy profile of cold dark matter (CDM). Consequently, solitons composed of ultra-light axions (ULAs) may resolve the `cusp-core' problem of CDM. In DM halos, thermodynamics will lead to a CDM-like Navarro-Frenk-White (NFW) profile at large radii, with a central soliton core at small radii. Using Monte-Carlo techniques to explore the possible density profiles of this form, a fit to stellar-kinematical data of dwarf spheroidal galaxies is performed. The data favour cores, and show no preference concerning the NFW part of the halo. In order for ULAs to resolve the cusp-core problem (without recourse to baryon feedback, or other astrophysical effects) the axion mass must satisfy $m_a<1.1\times 10^{-22} \text{ eV}$ at 95\% C.L. However, ULAs with $m_a\lesssim 1\times 10^{-22}\text{ eV}$ are in some tension with cosmological structure formation. An axion solution to the cusp-core problem thus makes novel predictions for future measurements of the epoch of reionisation. On the other hand, improved measurements of structure formation could soon impose a \emph{Catch 22} on axion/scalar field DM, similar to the case of warm DM.
\end{abstract}


\section{Introduction}

Dark matter (DM) is known to comprise the majority of the matter content of the universe \citep[e.g.][]{Ade:2013zuv,planck_2015_params}. The simplest and leading candidate is cold (C)DM. CDM has vanishing equation of state and sound speed, $w=c_s^2=0$, and clusters on all scales. Popular CDM candidates are $\mathcal{O}(\text{GeV})$ mass thermally produced supersymmetric weakly interacting massive particles \citep[SUSY WIMPs, e.g.][]{Jungman:1995df}, and the $\mathcal{O}(\mu\text{eV})$ mass non-thermally produced QCD axion \citep{pecceiquinn1977,weinberg1978,wilczek1978}. The free-streaming and decoupling lengths of a WIMP, and the Jeans scale of the QCD axion are both extremely small \citep[i.e. sub-solar on a mass scale, see e.g.][]{Loeb:2005pm}.

It is well known, however, that CDM faces a number of `small-scale' problems related to galaxy formation: `missing satellites' \citep{1999ApJ...524L..19M,1999ApJ...522...82K}, `too-big-to-fail' \citep{2011MNRAS.415L..40B}, and `cusp-core' \citep[reviewed in][]{2008IAUS..244...44W}, to name a few.  Baryonic physics, for example feedback or dynamical friction, offers possibilities to resolve some or all of these problems within the framework of CDM \citep[e.g.][and references therein]{2009ApJ...698.2093D,2012MNRAS.422.1231G,2014Natur.506..171P,2015arXiv150201947D}. Modifying the particle physics of DM so that it is no longer cold and collisionless also provides an attractive and competitive solution \citep[e.g.][]{Spergel:1999mh,HuBarkana&Gruzinov2000,2000ApJ...534L.127P,Boehm:2000gq,bode2001,marsh2013b,2014arXiv1412.1477E}. Modified gravity has also been considered in this context \citep[e.g.][]{Lombriser&Penarrubia2014}.

There is no compelling theoretical reason, however, that the DM should be cold or indeed thermal. From a pragmatic point of view, then, possible signatures of the particle nature of DM on galactic scales provide a useful tool to constrain or exclude models. The small-scale problems provide a useful way to frame our tests of DM using clustering. If the small-scale problems are resolved by baryon physics, then this removes one `side' of particle physics constraints derived from them. However, the other side is left intact: a particle physics solution cannot `over solve' the problem (e.g. too little structure or too large cores).

Thermal velocities and degeneracy pressure can support cores if the DM is warm \citep[WDM, e.g.][]{1982PhRvL..48.1636B,bode2001}. However, mass ranges allowed by large-scale structure constraints, $m_W\gtrsim \mathcal{O}(\text{few keV})$ limit core sizes to be too small \citep[e.g.][]{2012MNRAS.424.1105M,schneider2013b}. Particle physics candidates for WDM include sterile neutrinos, or the gravitino. Another promising solution is offered by ultra-light (pseudo) scalar field DM with $m_a\sim 10^{-22}\text{ eV}$ \citep{HuBarkana&Gruzinov2000,marsh2013b,2014NatPh..10..496S}, which may be in the form of axions arising in string theory \citep{witten2006,axiverse2009} or other extensions of the standard model of particle physics \citep[see][for a historic review]{Kim:1986ax}. In this model, cores are supported by quantum pressure arising from the axion de Broglie wavelength. Axions with $m_a> 10^{-23}\text{ eV}$ are consistent with large-scale structure constraints \citep{bozek2014} and can provide large cores \citep{HuBarkana&Gruzinov2000,marsh2013b,2014NatPh..10..496S,Schiveetal2014b}. Such ultra-light axions (ULAs) may therefore offer a viable particle physics solution to the small-scale problems. 

If the small-scale problems are in fact resolved by ULAs, then forthcoming experiments, such as ACTPol \citep{Calabrese:2014gwa}, the James Webb Space Telescope \citep[JWST,][]{2006NewAR..50..113W}, and {\em Euclid} \citep{Laureijs:2011gra,Amendola:2012ys}, will find novel signatures incompatible with CDM. These include a truncated and delayed reionisation history \citep{bozek2014}, a dearth of high-$z$ galaxies \citep{bozek2014}, and a lack of weak-lensing shear-power on small scales \citep[by analogy to WDM, e.g.][]{Smith:2011ev}. If these observations are consistent with CDM, however, then ULAs will be excluded from any relevant role in the small-scale crises.

We note that there is not consensus on the preference for cores versus cusps in dwarf density profiles \citep[e.g.][]{Breddels&Helmi2013,2014MNRAS.441.1584R,2014arXiv1406.6079S}. If dwarfs are in fact cuspy, then no baryon feedback or dynamical friction is necessary, and light DM providing cores would be excluded. In this work, we use the data of \cite{Walker&Penarrubia2011} (hereafter, WP11), who report that their measurements exclude NFW-like cuspy profiles at a confidence level of $95.9 \%$ for Fornax and $99.8\%$ for Sculptor.

The nature of axion clustering on small scales is a fascinating topic regardless of their role or otherwise in the small-scale crises. There has been much discussion in the literature \citep[e.g.][and references therein]{Sikivie:2009qn,Davidson:2013aba,Davidson:2014hfa,Guth:2014hsa,Banik:2015sma} concerning whether axions, including the QCD axion, undergo Bose-Einstein condensation and display long-range correlation. This question is of more than theoretical importance and can greatly affect the direct detection prospects for the axion, for example by ADMX \citep{Duffy:2006aa,Hoskins:2011iv}. The density solitons that we study here are the same ground-state solutions studied in the context of the QCD axion by \cite{Guth:2014hsa}, and display only short-range order. For ULAs, these solitons are kpc scale, while for the QCD axion they are closer to km scale.

DM composed of axions of any mass, be it the QCD axion or ULAs, possesses a characteristic scale and DM clustering should be granular on some scale \citep{2014NatPh..10..496S}. This scale is set by the axion Jeans scale, which varies with cosmic time, making structure formation non-hierarchical \citep{marsh2013b,bozek2014}. This departure from CDM on small scales makes the study of axion clustering on these scales impossible using standard N-body techniques, and we must ask many basic questions afresh. What fraction of the DM in the Milky Way is smooth, and what is in solitons? What is the power spectrum on small scales? What is the mass function of solitons and sub-halos? Answering these questions in detail will be the subject of future work, with the present work taking some small steps in this direction. The scale symmetry of the relevant equations for axion DM makes answers to these questions universal, and equally applicable to ULAs and the QCD axion.   

In this paper, we study the soliton solutions of axion DM, clarifying the core formation mechanism. The stability of this profile and its validity outside the spherically symmetric case is supported by pre-existing evidence from simulation. Our proposal for a complete density profile matching to NFW on large scales is phenomenological, and new to this work. We apply this to the dwarf spheroidal (dSph) galaxies, finding limits on the axion mass for a solution to the cusp-core problem. Our choice of data sets and methodology applied to the axion core model is entirely new to this work, and makes concrete links to cosmological limits. While we refer explicitly to our work as on `axion DM,' the results are more generally applicable to any non-thermal scalar DM candidate. We hope that the conclusions we draw inspire further study of these models within the community.

We begin in Section~\ref{sec:rapid_review} with discussion of some intuitive aspects of the soliton profile andthe relation between the central density, axion mass, and core radius. In Section~\ref{sec:halo_profile} we provide a complete model for the halo density profile for axion DM with explicit formulae. We use these formulae in Section~\ref{sec:Observations} to find limits on the axion mass based on stellar-kinematical data for Fornax and Sculptor dSph galaxies taken from WP11. We conclude in Section~\ref{sec:conclusions}. Derivations, pedagogical notes, and certain numerical aspects are relegated to the Appendix.

\section{Soliton Cores For Scalar DM}
\label{sec:rapid_review}

Quantum and wave-mechanical properties of axion DM allow for pressure support. Bose-Einstein condensation, if it occured, could also lead to large correlation lengths. Neither phenomenon will occur if the axion is modelled as pressureless dust with classical gravitational interactions. Although from the particle point of view, possible Bose condensation of the axion field is a quantum phenomenon, from the field point of view it can be studied classically \citep{Guth:2014hsa}. The classical analysis also allows one to derive the axion sound speed and resulting Jeans scale \citep{1985MNRAS.215..575K}. Indeed, the characteristic wavelength for the Bose-Einstein condensate, the scale over which thermalisation occurs and gravitational growth is suppressed, is none other than the Jeans scale in the fluid description of the classical field under linear density perturbations.

\subsection{Schr\"{o}dinger-Possion System and Ground State Solitons}

We work in the non-relativistic approximation, where the full Einstein-Klein-Gordon (EKG) equations reduce to the Schr\"{o}dinger-Poisson system \citep{Seidel:1990jh,Widrow&Kaiser1993}:
\begin{align}
i \hbar \frac{\partial \psi}{\partial t} &= -\frac{\hbar^2}{2m_a} \nabla^2 \psi +  m_aV \psi \, , \nonumber  \\
\nabla^2 V &= 4\pi G\psi \psi^* \, . 
 \label{eqn:SP_system}
\end{align}
The axion mass is $m_a$, and the field $\psi$ is related to the WKB amplitude of the axion field, $\phi$ (see Appendix~\ref{appendix:derivation}).

The soliton solutions of this system, and their relativistic completion, were first studied by \cite{1969PhRv..187.1767R}, who identified the scaling symmetry, and the maximum stable mass. High-resolution cosmological simulations of the Schr\"{o}dinger picture for DM by  \cite{2014NatPh..10..496S} using an adaptive mesh refinement scheme powered by GPU acceleration have provided evidence that ultra-light ($m_a\sim 10^{-22}\text{ eV}$) scalar DM halos form kpc scale soliton cores that are stable on cosmological time scales. 

Consider stationary wave solutions of the form:\footnote{Separating the function $\psi(r,t)$ into polar coordinates in this way is sometimes referred to as the Madelung transformation \citep{Madelung1926}. }
\begin{equation}
\psi(r,t) = e^{-i \gamma t} \chi(r)
\label{eqn:soliton_ansatz}
\end{equation}
for a function $\chi(r)$ which depends on the radial coordinate $r$, but does not vary with time. Taking $\partial_r \gamma=0$ assumes phase coherence and zero fluid velocity. We expect loss of phase coherence at large distances in real systems, and expect that this is related to the transition from soliton to NFW profile discussed in Section~\ref{sec:halo_profile}. Indeed, phase coherence is lost outside of the density cores in the simulations of \cite{2014NatPh..10..496S}. 

In spherical co-ordinates we have the following system of ODEs:
\begin{align}
\chi'' + \frac{2\chi'}{r} &= 2(V-\gamma)\chi \, , \nonumber \\
V'' + \frac{2V'}{r} &= \chi^2 \, , \label{eqn:SP_System2}
\end{align}
where $V$, $\chi$, $r$, and $\gamma$ are all dimensionless quantities defined in Appendix~\ref{appendix:dimensions}. The solutions $\chi(r)$ are a special type of soliton known as an oscillaton (see Appendix~\ref{appendix:soliton_note} for clarification of terminology) in the axion field $\phi$, with soliton profile $\chi$, and density profile $\rho=\chi^2$. 

In the stationary wave solution, Eq.~(\ref{eqn:soliton_ansatz}), it is important to note that $\gamma$ is a parameter to be solved for, related to the total energy. The ground state is the state of lowest energy and depends on only one length scale. It also has no nodes, and with fixed central density this allows us to find the unique value of $\gamma$ numerically. The stationary wave solution must obey the condition $|m\psi|\gg |\hbar\dot{\psi}|$ (first order WKB approximation: see Appendix~\ref{appendix:derivation}), which translates to $\gamma\ll 1$ in dimensionless variables. This must be satisfied if solutions to Eqs.~\ref{eqn:SP_System2} are consistent non-relativistic limits of the solutions to the full EKG equations. 

The soliton density profile must have the properties $\tilde{\rho}(0)=\text{const.}$ and $\tilde{\rho}(r\rightarrow\infty)=0$, such that $\tilde{\rho}$ is the solution of the boundary value problem in Appendix~\ref{sec:schrodinger_numerical}.We write the density profile as:
\begin{equation}
\tilde{\rho}_{\rm sol}(r)=f(\alpha r) \, ,
\end{equation}
where $f(y)$ has no explicit scales and has the correct asymptotic behaviour. Restoring units:
\begin{align}
\rho_{\rm sol}(r)&=2m_a^2M_{pl}^2 f( r/r_{\rm sol}) \, , \\
r_{\rm sol}&:= \frac{1}{\alpha m_a} \, .
\end{align}

Solutions to Eqs.~\ref{eqn:SP_System2} possess a scaling symmetry \citep{1969PhRv..187.1767R}, as discussed in Appendix~\ref{appendix:scaling}, which is used to fix the appropriate scale for astrophysical systems. Upon applying a rescaling by $\lambda$, all dimensionful quantities\footnote{Except the axion mass, which does not appear explicitly in Eq.~(\ref{eqn:SP_System2}), and hence only sets the units.
are affected, so that $r\rightarrow r/\lambda$ and $r_{\rm sol}\rightarrow r_{\rm sol}/\lambda$.} The density must scale as $\rho\rightarrow \lambda^4\rho$. Rewriting
\begin{equation}
\frac{\rho_{\rm sol}(r)}{\rho_{\rm crit}} = \delta_{\rm sol} f(r/r_{\rm sol}) \, ,
\label{eqn:def_f_delta}
\end{equation}
we find the relationship between the soliton density parameter, $\delta_{\rm sol}$, and its characteristic radius, $r_{\rm sol}$:
\begin{equation}
\delta_{s} = \left(\frac{5\times 10^4}{\alpha^4} \right) \left(\frac{h}{0.7}\right)^{-2}\left(\frac{m_a}{10^{-22}\text{ eV}}\right)^{-2}\left(\frac{r_{\rm sol}}{\text{kpc}}\right)^{-4} \, .
\label{eqn:delta_sol_r_sol}
\end{equation}
The value of $\alpha$ is fixed from numerical fits in Appendix~\ref{appendix:fits}.

A rescaling of the density profile is seen to affect only $\delta_{\rm sol}\rightarrow \lambda^4 \delta_{\rm sol}$, with the relationship between the central density, the scale radius, and the mass completely being fixed. The numerical value of $\alpha$ depends on the choice of the functional form of $f(\alpha r)$. Choosing a different definition for $r_{\rm sol}$, e.g. the half-density radius used by \cite{2014NatPh..10..496S,Schiveetal2014b}, or even a completely different functional form for $f$ will only change the numerical co-efficient in Eq.~(\ref{eqn:delta_sol_r_sol}) and not the functional relationship between $\delta_{\rm sol}$ and $r_{\rm sol}$, or their dependence on the axion mass. These features are universal.  In Appendix~\ref{sec:schrodinger_numerical}, we fit the form of $f(\alpha r)$ from numerical solutions, and fix the value of $\alpha$ for our chosen fit.

The axion mass can be re-expressed in terms of the linear Jeans scale, $r_{J,{\rm lin}}/\text{kpc}\propto (m_a/10^{-22}\text{ eV})^{-1/2}$ \citep{HuBarkana&Gruzinov2000}. Substituting into Eqs.~(\ref{eqn:def_f_delta},\ref{eqn:delta_sol_r_sol}), the scale radius of any soliton with central density $\rho_{\rm sol}(0)$ is given by
\begin{equation}
r_{\rm sol} \propto \left( \frac{\rho_{\rm sol}(0)}{\rho_{\rm crit}}\right)^{-1/4} r_{J,{\rm lin}} \, .
\label{eqn:rsol_scaling}
\end{equation}
This simple result fixes the scaling properties of axion density cores using only the scaling symmetry of the SP system. The existence of the scale in the ground state was guaranteed by the uncertainty principle, which holds by virtue of the large occupation numbers and the classical wave-mechanical description of the axion field. The scaling symmetry occurs because in the non-relativistic limit there is no scale in the SP system. Such a scaling should therefore hold for all large occupation number, non-relativistic axion/scalar DM density configurations in the small-radius and long-time limits. That is, \emph{soliton cores with this scaling property are universal features} of such DM models.

\subsection{Soliton Core Radius}
\label{sec:uncertainty}
\begin{figure}
\centerline{\includegraphics[width = .5\textwidth]{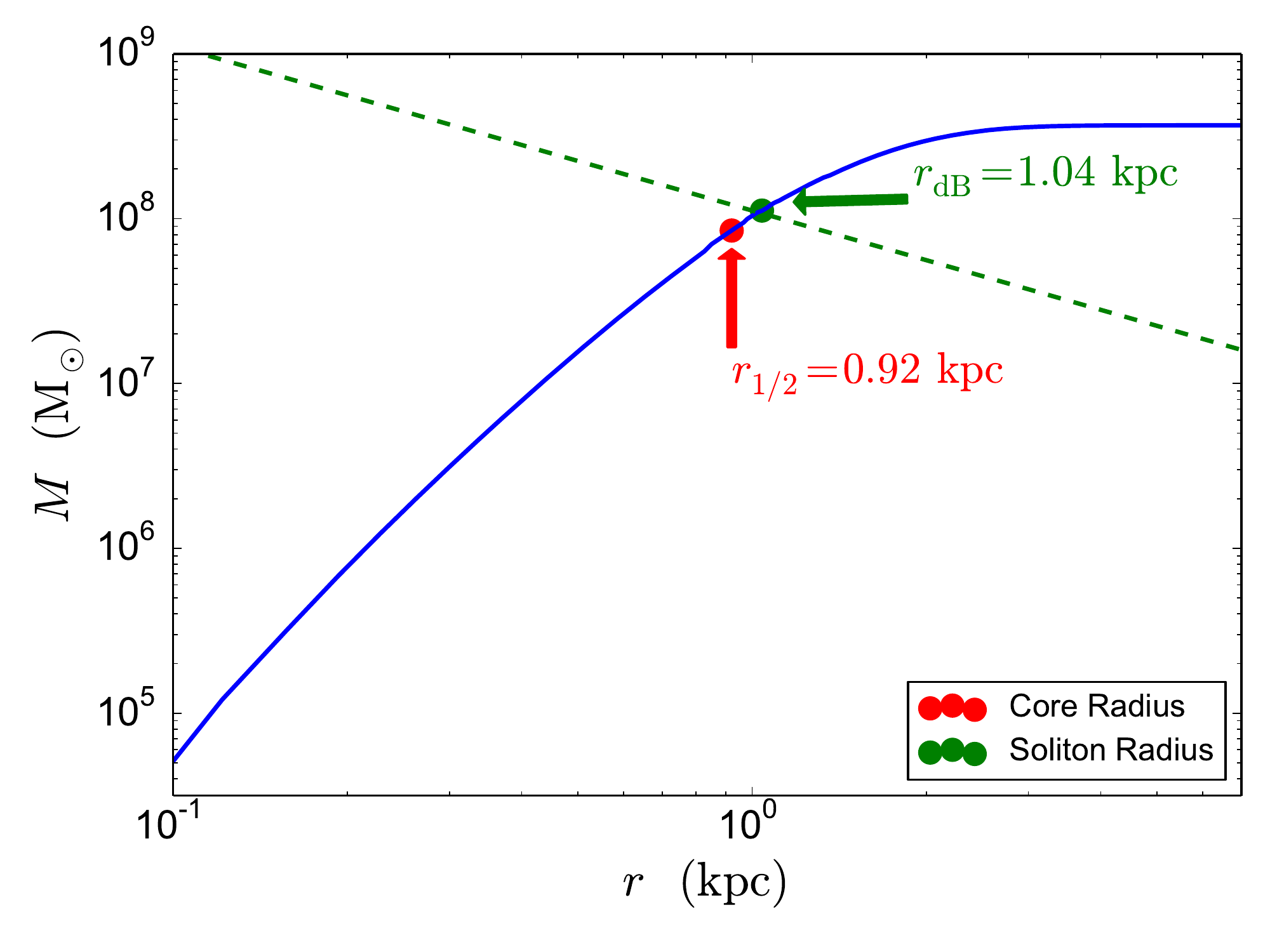}}
\caption{Solving Eq.~(\ref{eqn:debroglie}). Blue: left hand side, the mass of the soliton as a function of radius for some representative dwarf galaxy parameters. Green: right hand side. The intersection solves for the de Broglie radius thus defined. The red dot shows the core radius as defined by \protect\cite{2014NatPh..10..496S}, $r_{1/2}$, while the green dot shows the de Broglie radius obtained from the uncertainty principle. Due to the steeply falling soliton profile outside of the core, neither definition of core radius fixes the density of the point of transition to NFW accurately \citep{2014NatPh..10..496S,Schiveetal2014b}.} \label{fig:SolitonMass}
\end{figure}

We now consider the relation of the soliton radius to the de Broglie wavelength. Consider a particle of mass $m_a$ on a circular orbit around the soliton core. Its velocity is given by:
\begin{equation}
v^2 = \frac{GM(<r)}{r}  \label{eqn:velocityCircular}
\end{equation}
The uncertainty principle requires:
\begin{equation}
p \, r \geq \hbar
\end{equation}
The de Broglie scale is the point where $p\, r_{\rm dB}  = \hbar$, and thus:
\begin{equation}
r_{\rm dB} m_a \sqrt{\frac{G	M(<r_{\rm dB})}{r_{\rm dB}}} = \hbar
\end{equation}
\begin{equation}
\Rightarrow M(<r_{\rm dB}) = \frac{\hbar^2}{m_a^2 G}\frac{1}{r_{\rm dB}} \, .
\label{eqn:debroglie}
\end{equation}
In Fig.~\ref{fig:SolitonMass} we solve this equation graphically for some typical dwarf galaxy parameters and find $r_{\rm dB} = 1.04 \,\mathrm{kpc}$. The de Broglie wavelength found in this way is close to the core radius at half central density, $r_{1/2}$, where $\rho_{\rm sol}(r_{1/2}) = \rho_{\rm sol}(0)/2$. 

The form of the scaling in Eq.~(\ref{eqn:rsol_scaling}) is the same as for the halo Jeans scale extrapolated from linear theory \citep{HuBarkana&Gruzinov2000,marsh2013b,Guth:2014hsa}. Both follow from the scale symmetry and dimensional analysis, and thus their interpretation as related to the de Broglie scale remains the same \citep[see also][]{Hlozek:2014lca}. Beyond scaling alone, how exactly are the linear Jeans scale and the soliton size related numerically? Can this be used to predict properties of axion halos, including the core-size/mass relation, and the transition to NFW? 
\begin{table*}
\centering
\begin{tabular}{r|r|r|r|r}
\hline \hline
dSph&$\log_{10}(\sigma^2/$km$^2$ s$^{-2})$& Error & $\log_{10}(r/$kpc) & Error  \\ \hline \hline
Fornax&2.00 & 0.05 & -0.26 & 0.04 \\ 
& 2.32 & 0.04 & -0.05& 0.04 \\ \hline
Sculptor&1.62 & 0.06 & -0.78 & 0.04 \\ 
& 2.13 & 0.05 & -0.52& 0.04 \\ \hline\hline
\end{tabular}
\caption{Data used in this work, taken from WP11. We approximate the likelihoods to be two-dimensional uncorrelated Gaussians in $\log_{10}(r,\sigma)$ for each data point.}
 \label{tab:walker_data}
\end{table*}

The simulation results of \cite{2014NatPh..10..496S} indicate a soliton core connecting to an NFW profile at larger radii. No prescription for the matching radius is provided, however. In a follow up paper \citep{Schiveetal2014b}, the same authors assessed halo formation in this model as the gravitational collapse of a system of solitons. Using the uncertainty principle and the virial theorem they motivated a relationship between the total halo mass, $M_h$, the soliton core mass $M(<r_{1/2})$, and the minimum halo mass, $M_{\rm min}$: $M(<r_{1/2})\propto (M_h/M_{\rm min})^{1/3}M_{\rm min}$.

\cite{HuBarkana&Gruzinov2000} and \cite{marsh2013b} related the core radius to the halo Jeans scale. Can the same quantity be used to define a matching point between soliton and NFW? With respect to the linear Jeans scale $r_J$, the halo Jeans scale, $r_{J,h}$, satisfies
\begin{equation}
\frac{\rho(r_{J,h})}{\rho_{\rm crit}}=\left(\frac{r_{J}}{r_{J,h}}\right)^4 \, .
\label{eqn:rjh_extrap}
\end{equation}

In order for this relation to define a matching point between soliton and NFW profiles, both profiles must provide a solution to this equation at the same radius. The logarithmic slope, $\Gamma$, of the NFW profile is in the range $-3<\Gamma<-1$ and so there is always a single unique solution for $r_{J,h}$. However, the slope of the soliton profile is zero at the origin, decreasing to $\Gamma<-4$ at large radius. For the soliton, there can be either zero, one or two solutions for $r_{J,h}$, with a single unique solution at the point where $\Gamma=-4$. Using the relationship in Eq.~(\ref{eqn:delta_sol_r_sol}) it is possible to show that solutions only occur for $\delta_s \ll 1$. Halos, on the other hand, correspond to non-linear density perturbations with $\delta_s\gg 1$. Therefore, the transition from soliton to NFW profile in a halo must occur on radii smaller than the linear Jeans scale extrapolated using the local density, i.e. on $r<r_{J,h}$ defined by Eq.~(\ref{eqn:rjh_extrap}). This implies that \emph{soliton cores are more compact than expected from linear theory}. This will turn out to have important implications for a possible solution to the cusp-core problem using ULAs/scalar field DM.

\section{Axion Halo Density Profiles and Dwarf Galaxy Cores}

In this section we will first define our proposal for the complete halo density profile of axion /scalar-field DM. We then go on to estimate the parameters in this profile using the Fornax and Sculptor dSph density profile slopes as measured by WP11. We introduce this measurement and the approximations it uses, define a likelihood from this, and perform a Monte Carlo Markov chain (MCMC) analysis. The results are all contained in Table~\ref{tab:results} and Figs.~\ref{fig:hudf_triangle} and \ref{fig:samples_panel}.

\subsection{Defining the density profile}
\label{sec:halo_profile}

The halo density profiles for ultra light scalar dark matter observed in the simulations of \cite{2014NatPh..10..496S,Schiveetal2014b} consist of an NFW-like outer region, with a prominent solitonic core. The observed transition is sharp, and we model it as a step function:
\begin{equation}
\rho(r)=\Theta (r_\epsilon-r)\rho_{\rm sol}(r)+\Theta (r-r_\epsilon)\rho_{\rm NFW}(r) \, . \label{eqn:total_profile}
\end{equation}

The transition to an NFW profile at some radius is expected on a number of physical grounds. Axion/scalar field DM is indistinguishable from CDM on large scales, and this is the basis of the Schr\"{o}dinger approach to simulating CDM \citep{Widrow&Kaiser1993,Uhlemann:2014npa}. Thus, on scales much larger than the de Broglie wavelength, the scalar field halos should resemble those found in N-body simulations, i.e. NFW. Furthermore, no long range correlation should occur; the smallest objects are solitons, with a characteristic granularity fixed by the Jeans scale \citep{2014NatPh..10..496S,Schiveetal2014b,Guth:2014hsa}. On large scales, phase decoherence should occur, violating the assumption in our soliton ansatz that $\partial_r \gamma=0$. 

The dynamics will be that of an interacting gas of solitons in a decoherent scalar field background. On length scales larger than the soliton radius the usual thermodynamic arguments relating to dust \citep{2008gady.book.....B} will apply, leading to an NFW-like profile \citep[on large scales this equivalence between the Schr\"odinger picture and dust thermodynamics can be derived using the Wigner distribution,e.g.][]{Widrow&Kaiser1993,Uhlemann:2014npa}. The transition in behaviour from soliton to NFW governed by the decoherence scale, and the mass function of solitons in the outer halo are all interesting questions, and will be the subjects of forthcoming papers.	
The NFW density profile is given by \citep{Navarroetal1997}
\begin{equation}
\frac{\rho_{\rm NFW}(r)}{\rho_{\rm crit}}=\frac{\delta_{\rm char}}{(r/r_s)(1+r/r_s)^2} \, .
\end{equation}
We fit for the soliton density profile in Appendix~\ref{appendix:fits}
\begin{equation}
\frac{\rho_{\rm sol}(r)}{\rho_{\rm crit}} = \frac{\delta_s}{(1+(r/r_{\rm sol})^2)^8} \, .
\end{equation}
We take $\delta_s$ as a free parameter and $r_{\rm sol}$ is fixed in terms of it by Eq.~(\ref{eqn:delta_sol_r_sol}) using $\alpha=0.230$. The rescaling parameter, $\lambda$, can then be found by setting $r_{\rm sol}=(\lambda \alpha m_a)^{-1}$. We recall that consistent solutions with $|\gamma/m|\ll 1$ require $\lambda<1$, which also guarantees $\phi(0)\lesssim 0.3 M_{pl}$ and the stability of the soliton \citep{Seidel:1990jh}.

We match the soliton and NFW profiles at a fixed value of the overdensity, $\delta=\epsilon\delta_s$, defining the matching radius $r_{\epsilon}$. Since we have not been able to find an accurate estimate that fixes $\epsilon$ analytically, we take it as a free parameter.\footnote{An order of magnitude estimate for the matching radius based on the de Broglie scale will not be enough. The soliton density falls rapidly for $r>r_{\rm dB}$ and so $\mathcal{O}(1)$ numerical co-efficients have a large effect on the estimate for $\epsilon$.} Continuity of the density at this point fixes one of the two free parameters in the NFW profile. We choose this to be $\delta_{\rm char}$ and take the scale radius to be a free parameter. For a given ULA mass, our halo density profile thus has three additional free parameters:
\begin{equation}
\{\delta_s,\epsilon,r_s\} \, .
\end{equation}

We emphasise again that a theoretical model for the value of $\epsilon$, which could depend on redshift, central density and/or particle mass, could perhaps be derived. In this case, the density profile has just as many free parameters as an NFW profile. Thus, in any given halo, a core measurement would predict the point of transition to NFW, and a measurement of the outer halo fixing concentration and scale radius would predict the corresponding inner core size.

The data we use imply cored profiles on the observed radii, and so in our phenomenological model with free $\epsilon$ we will find that  the NFW parameters $\epsilon$ and $r_s$ are unconstrained. This shows that the data prefer soliton cores to NFW profiles. For a complete Bayesian analysis, we include and marginalise over the NFW parameters in our constraints, as described in the next subsection.

\subsection{Fitting to Fornax and Sculptor}
\label{sec:Observations}

\begin{table}
\centering
\begin{tabular}{r|r}
\hline \hline
Parameter&Prior  \\ \hline \hline
$\log_{10}(m_a/\text{eV})$&$U(X,-19)$ \\
$\log_{10}\delta_s^{\rm F,S}$&$U(0,10)$ \\
$\log_{10}\epsilon^{\rm F,S}$&$U(-5,\log_{10}0.5)$ \\
$\log_{10}(r_s^{\rm F,S}/\text{kpc})$&$U(-1,2)$ \\ \hline\hline
\end{tabular}
\caption{Priors on our density profile parameters, where $U(a,b)$ is the uniform distribution on $[a,b]$. The lower bound on the mass prior, $X$ is given two different values: $X_{\rm CMB}=-25$ \citep{Hlozek:2014lca} and $X_{\rm HUDF}=-23$ \citep{bozek2014}, referred to as CMB and HUDF priors respectively.}
 \label{tab:priors}
\end{table}
\begin{table}
\centering
\begin{tabular}{r|r}
\hline \hline
Parameter&Posterior  \\ \hline \hline
$(m_a/\text{eV})_{\rm HUDF}$&$<1.1 \times 10^{-22}$  (95\% C.L.) \\
$(m_a/\text{eV})_{\rm CMB}$&$<1.0 \times 10^{-22}$  (95\% C.L.) \\
$\log_{10}\delta_s^{\rm F}$ & $5.55^{+0.06}_{-0.08}$\\
$\log_{10}\delta_s^{\rm S}$ & $6.19\pm 0.05$\\ \hline\hline
\end{tabular}
\caption{Posteriors on the constrained density profile parameters. The mass constraint is quoted for both the CMB and HUDF priors (see Table~\ref{tab:priors}). Upper and lower errors on the central densities are given as the 16th and 84th percentiles and are quoted for the HUDF prior only (see text for discussion). The upper bound on the axion mass reflects the minimum core size consistent with observations.}
 \label{tab:results}
\end{table}	

While there are several ways of analysing the observed data, we will focus on the method used by WP11 who measured the slopes of dSph mass profiles directly from stellar spectroscopic data. They use the fact that some dSphs have been shown to have at least two stellar populations that are chemo-dynamically distinct \citep[see e.g.][]{Tolstoyetal2004,Battagliaetal2006,Battagliaetal2011}. Measuring the halflight radii and velocity dispersions of two such populations in a dSph allows one to infer the slope of the mass profile. The method of WP11 has the advantage that it does not need to adopt any dark matter halo model a priori. Therefore the data can be used to test theoretical density profile models without having to run computationally expensive fits to the full stellar data for each value of the theoretical parameters. 
\begin{figure*}
\vspace{-0.2em}\includegraphics[width=1.9\columnwidth]{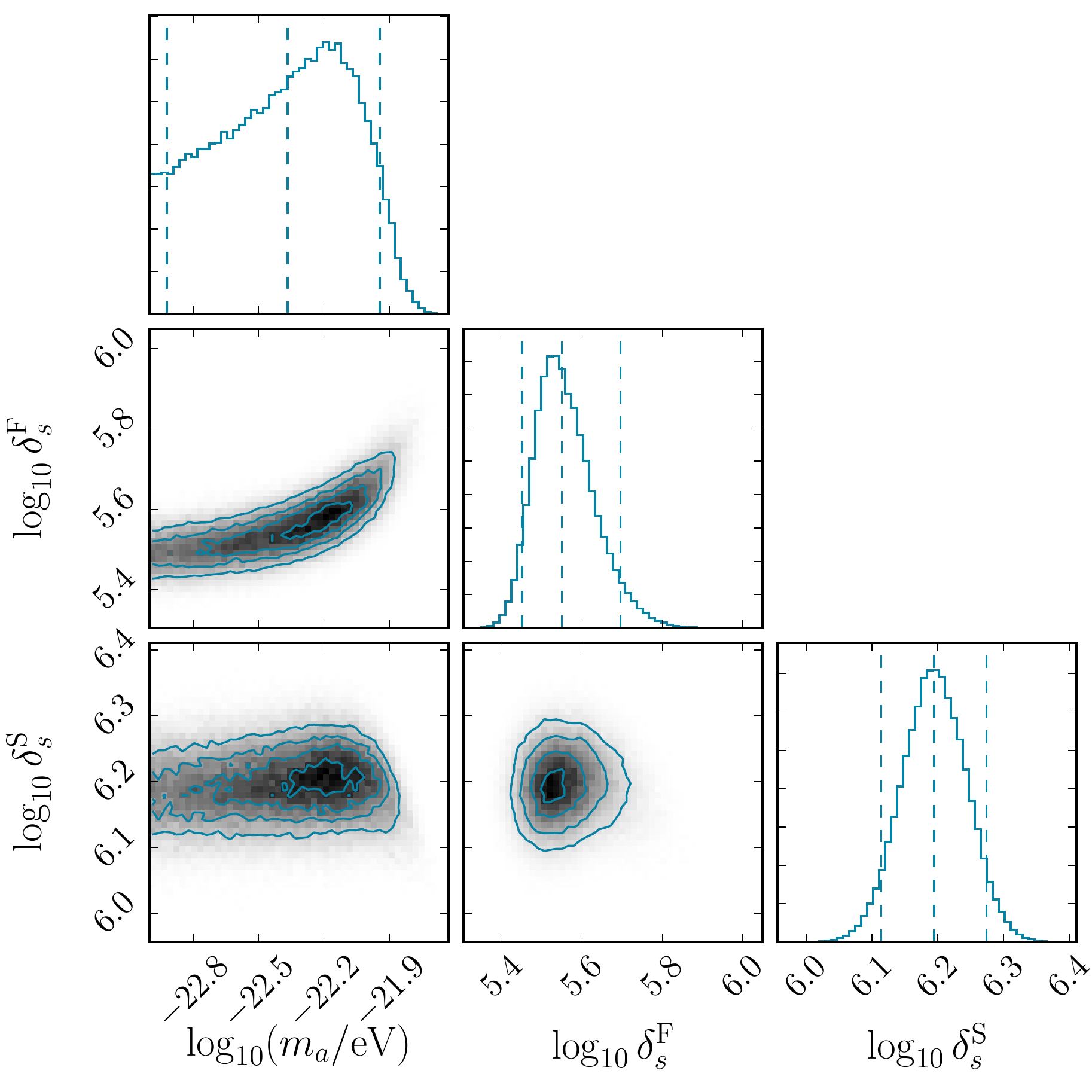}
\caption{Two and one-dimensional marginalised posteriors on the constrained density profile parameters for the HUDF mass prior (see Table~\ref{tab:priors}). Contour levels show $[0.5,1.0,1.5,2.0]\sigma$ preferred regions. Vertical dashed lines show $[0.05,0.5,0.95]$ percentiles. This plot is made using \textsc{triangleplot} \citep{triangle_plot}.}
\label{fig:hudf_triangle}
\end{figure*}	

We model the results of WP11 as providing two-dimensional Gaussian distributions for the half light radii, $r_{h,i}$, and velocity dispersions, $\sigma(r_{h,i})$, for each of the two stellar populations, $i$, in Fornax and Sculptor. While the exact results display some covariance between $\sigma$ and $r_h$ the Gaussian approximation is far simpler to analyse, and accurate enough for the purposes of this study. This follows the approach taken by \cite{Lombriser&Penarrubia2014} for testing chameleon gravity using dSphs. The data we use are given in Table~\ref{tab:walker_data}.

For our density profile the mass internal to any radius, $M(<r)$, can be computed analytically (and so can the derivative of the mass, $dM/dr$). These are the only ingredients necessary in anlaysing this simplified version of the stellar kinematic data. The velocity dispersion at the stellar half-light radius obeys the following empirical relationship:
\begin{equation}
\sigma^2(r_h)=\frac{2 G M(<r_h)}{5 r_h} \, .
\label{eqn:empirical_sigma}
\end{equation}
This relationship is related to the virial theorem, and is found by solving the Jeans equation and finding a `sweet spot' where a wide variety of density and velocity profiles agree (including projection and anisotropy effects). In this work we do not perform a full Jeans analysis and use only this analytic relationship and the derived errors on it, following WP11. 
\begin{figure*}
\vspace{-0.2em}\includegraphics[width=2\columnwidth]{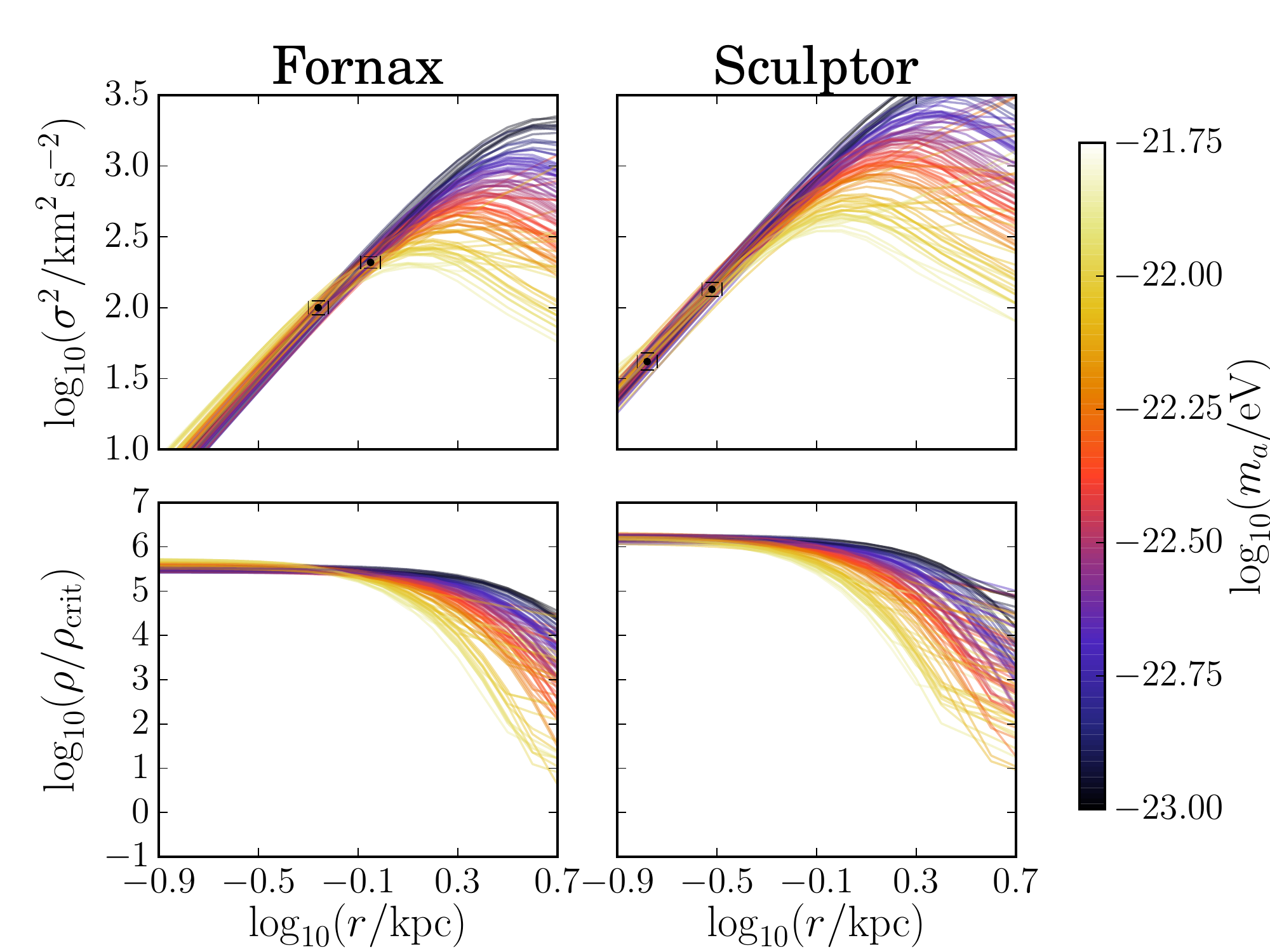}
\caption{Velocity dispersions are estimated at the stellar half-light radii from $M(<r_h)$, Eq.~(\ref{eqn:empirical_sigma}). We show velocity dispersions and density profiles of 200 random samples from our MCMC using the HUDF mass prior (see Table~\ref{tab:priors}). The large core in Fornax favours low axion masses. Consistency with HUDF predicts a rapid turnover in the density slope at radii slightly larger than those observed, caused by the transition from soliton to NFW. Data from WP11, quoted in Table~\ref{tab:walker_data}.}
\label{fig:samples_panel}
\end{figure*}	

The two-dimensional nature of the $(r,\sigma)$ data can be accounted for using an effective one-dimensional error \citep[e.g.][]{2013ApJ...771..137M}. If the data has central values $(\bar{x},\bar{y})$ and standard deviations $(\Sigma_x,\Sigma_y)$ then for a given model $y=f(x)$ the effective one-dimensional error is given by:
\begin{equation}
\Sigma_{\rm eff}^2=\Sigma_y^2+\left(\frac{df(\bar{x})}{dx}\right)^2\Sigma_x^2  \, .
\end{equation}
The likelihood, $\mathcal{L}$, can then be approximated as
\begin{equation}
\mathcal{L} \propto \exp \left[\frac{-(f(\bar{x})-\bar{y})^2}{2\Sigma_{\rm eff}^2} \right] \, .
\end{equation}

We use the data for Fornax and Sculptor with equal weight in the combined likelihood. The axion mass, $m_a$, is a global parameter which is the same for both Fornax and Sculptor. Our final model thus has seven parameters in total:
\begin{equation}
\{m_a,\delta_{s}^{\rm F},\delta_s^{\rm S},r_s^{\rm F},r_s^{\rm S},\epsilon^{\rm S},\epsilon^{\rm F}\} \, ,
\end{equation}
with F, S, labelling Fornax and Sculptor respectively. 

The priors on these parameters are given in Table~\ref{tab:priors}. We take Jeffreys', or `least information', priors in a fixed range for all parameters. We will discuss our results in detail shortly, but mention here those aspects relevant to priors. We find that the NFW parameters $(\epsilon, r_s)$ are unconstrained. The priors on these parameters are therefore irrelevant in quoting marginalised constraints on other parameters and we take them extremely wide to explore all possibilities. We impose an upper bound on the $\epsilon$ prior of $\epsilon=0.5$, so that the match must occur outside the half-density radius, consistent with the simulation results of \cite{2014NatPh..10..496S}. As already mentioned, the fact that these parameters are unconstrained has physical meaning: our MCMC finds no peak corresponding to an NFW profile, and thus the data prefer soliton cores to NFW cusps. The central densities are well constrained and so the results are independent of the prior. 

We find a one-sided constraint on the axion mass, and so the percentiles quoted depend on the prior for the lower bound. We take two such priors. The first uses the results of \cite{Hlozek:2014lca}, which place an approximate lower bound on ULAs to be all of the DM of $m_a>10^{-25}\text{ eV}$. This is a very conservative lower bound and relies only on linear constraints from the CMB. It is thus extremely reliable. Our alternative prior uses the results \cite{bozek2014}, which constrains $m_a>10^{-23}\text{ eV}$ at more than $8\sigma$ significance using Hubble Ultra-Deep-Field (HUDF). While this is a very strong bound, it relies on more assumptions about the astrophysics of reionisation and on the non-linear structure formation of ULAs. This HUDF bound is, however, still rather conservative compared to other constraints to ULAs using non-linear scales \citep[e.g. Lyman-alpha forest,][]{Amendola:2005ad}. 

We evaluate the likelihood using \textsc{emcee}, an affine-invariant MCMC ensemble sampler \citep{emcee}. We use 200 `walkers.' Convergence is tested by evaluating the auto-correlation time, which is found to be $\mathcal{O}(80)$ for each parameter. With 5000 steps per walker the likelihood is thus evaluated over many correlation times, giving many independent samples. We have used an `MCMC hammer' \citep{emcee} to crack a very simple parameter-constraint-nut, but this gives us a high degree of confidence in our results, and allows us to present them in a completely Bayesian fashion.

Our results are presented in Table~\ref{tab:results}, and shown for the HUDF priors in Fig.~\ref{fig:hudf_triangle}. As already mentioned, the NFW parameters $(r_s,\epsilon)$ are unconstrained, and so we do not show them. A weak constraint could be found on these parameters if one were to solve the full Jeans equation to fit the stellar kinematic data, rather than using only the empirical relationship Eq.~(\ref{eqn:empirical_sigma}). The constraint is caused by requiring a fixed enclosed mass at large radius. This consequently also imposes a weak lower bound on the axion mass to give a finite core radius, as in \cite{2014NatPh..10..496S}. Such an analysis is beyond the scope of this paper. Information about the total halo virial mass could also allow one to impose the core-halo relation of \cite{Schiveetal2014b}, introducing a degeneracy between $r_s$ and $\epsilon$.

For the central densities, shifts in the central values using the CMB prior compared to the HUDF prior are smaller than $1\sigma$ for Fornax and Sculptor, with comparable errors in both cases. The shift for Fornax is ever-so-slightly larger due to the degeneracy between $\delta_s$ and $m_a$ enforced by Eq.~(\ref{eqn:delta_sol_r_sol}) having an effect when marginalising over $m_a$ with a different prior in the unconstrained, low mass region. The degeneracy between these parameters is more pronounced for Fornax as it is less cored than Sculptor. This same effect is the cause of the non-Gaussianity in the one-dimensional central density distribution for Fornax.   

Random samples from our \textsc{emcee} runs showing the density and velocity profiles colour-coded by axion mass are shown in Fig.~\ref{fig:samples_panel} for the HUDF prior. This makes one clear point: since the HUDF prior already restricts the mass, cores cannot be too large. Therefore, if ULAs are responsible for dSph cores, velocity dispersions and density profiles must turn over quite rapidly on radii just larger than those observed. This is, in principle, a falsifiable prediction of the ULA model for cores.
\vspace{-0.1in}
\section{Discussion and Conclusions}
\label{sec:conclusions}

In this paper we have investigated stable solitonic density profiles, which are expected to be the smallest structures formed in models of axion/scalar field  DM. We found these objects via numerical solution of the boundary value problem in the Schr\"odinger approach to DM. Via use of a scaling symmetry we found a universal relationship between the central density of the soliton, the soliton radius and the DM particle mass. We further explained the soliton size by recourse to the uncertainty principle. The standard thermodynamic arguments for dust, and the formal equivalence between the Schr\"odinger picture and dust on large scales, further supported by evidence from simulation, lead us to consider solitons as the central cores in NFW halos. We presented a new phenomenological model for such a density profile.

If the DM is ultra-light, then solitons may be responsible for the density cores observed in dSph galaxies, with soliton density profiles in the inner regions, and an outer NFW profile. We investigated the validity of this claim by performing an MCMC analysis of the simplified stellar kinematic data of WP11 \citep{Walker&Penarrubia2011} for Fornax and Sculptor. This data shows a preference for cores over cusps, but other studies prefer cusps \citep[e.g.][]{Breddels&Helmi2013,2014MNRAS.441.1584R}. The preference of the WP11 data for large cores shows up as an upper bound on the axion DM mass of $m_a<1.1\times 10^{-22}\text{ eV}$ at 95\% C.L., leaving the NFW parameters unconstrained and marginalised over. The axion mass bound applies only if the dSph cores are solitonic, and not if they are caused by baryonic effects for standard CDM (be it composed of heavier axions, WIMPs etc.). This bound is consistent with the best fit mass $m_a=8.1^{+1.6}_{-1.7}\times 10^{-23}\text{ eV}$ (an approximate 95\% C.L. region $0.47\lesssim m_a/10^{-22}\text{ eV}\lesssim 1.13$) of \cite{2014NatPh..10..496S} found by solving a simplified version of the full Jeans equations for a single stellar sub-component in Fornax alone. \cite{2014NatPh..10..496S} also checked the (rough) consistency of their results with density profile slope measurements, but our analysis is the first to use these data together in a complete and Bayesian manner. In future we hope to conduct a full Jeans analysis for our model, following the similar analysis for self-interacting scalar field DM by \cite{2014PhRvD..90d3517D}.

The existence of an upper bound on the DM particle mass in the axion model has a number of interesting consequences. Structure formation in this model is substantially different from CDM due to the cut-off in the power spectrum on small scales caused by the axion sound speed and resulting Jeans scale. \cite{bozek2014} found that, if the DM is composed entirely of ultra-light axions, $m_a=10^{-22}\text{ eV}$ is \emph{just} consistent with HUDF. Our upper bound is right on the edge of this limit, and suggests two possible outcomes warranting further study:
\begin{itemize}
\item {\bf Pessimistic: }\emph{Structure formation and the requirement of dSph cores put conflicting demands on axion DM. This places the model in a `Catch 22' analogous to WDM.\footnote{Phrase coined by \cite{2012MNRAS.424.1105M}. For ULAs, one might call it a \emph{Catch} $10^{-22}$.} This particle physics model is no longer a catch-all solution to the small-scale crises, and additional mechanisms are required for its consistency.}
\item {\bf Optimistic: }\emph{Ultra-light axions are responsible for dSph cores. The cut-off in the power spectrum is just outside current observational reach. Near-future experiments will turn up striking evidence for axions in structure formation and in study of the high-$z$ universe.} 
\end{itemize}

In the pessimistic case, axions have nothing to do with core formation in dSph galaxies, and another mechanism is needed to form the cores. The DM can still be composed of axions, but there is now no limit on the mass based on the dSph observations. Cores may be formed by baryonic processes, putting axion DM in the same situation as any other DM model, e.g. WIMPs. Cores could also be formed by adding strong self-interactions for CDM or for axions. One could also modify gravity, for example if the modulus partner of the axion played the role of a chameleon field \citep{Khoury:2003rn,Lombriser&Penarrubia2014}. Another alternative would be to keep cores from axion solitons, but compensate structure formation by a boost in the primordial power or some other alteration to cosmology. A compensation based on a mixed DM model, however, is unlikely to succeed \citep{marsh2013b}.

There are still opportunities to discover evidence for axion DM in the pessimistic case. We briefly outline some of these for the `most pessimistic' case (from a particle physics view point) that cores are formed entirely by baryonic processes, and consider only searches influenced in some way by the Jeans scale or solitons, i.e. by non-trivial gravitational behaviour. For $10^{-22}\text{ eV}\lesssim m_a \lesssim 10^{-20}\text{ eV}$ axions are consistent with current constraints from cosmology, but the cut off in the power spectrum may still be observable using measurements of the weak lensing or 21cm power spectra. Axions can further be evidenced by their effect on black holes via the superradiant instability \citep{Arvanitaki:2010sy}. Spinning super-massive black holes indirectly probe/constrain masses in the range $10^{-19}\text{ eV}\lesssim m_a \lesssim 10^{-18}\text{ eV}$ \citep{Pani:2012vp,Brito:2014wla}. More direct signatures in gravitational wave detection may come from solar mass black holes for axions in the range $10^{-13}\text{ eV}\lesssim m_a \lesssim 10^{-10}\text{ eV}$ \citep{Arvanitaki:2014wva}. 

Axion DM of any mass will contain solitons somewhere in the mass spectrum at some point in cosmic history. This could have a number of consequences relevant to both the optimistic and pessimistic scenarios outlined above. This model predicts small and dense cores in all DM halos, with core size inversely related to central density and halo mass \citep[Eq.~\ref{eqn:delta_sol_r_sol}, and][]{Schiveetal2014b}. Such cores may provide seeds for high-$z$ quasars \citep{Schiveetal2014b}. The granular nature of DM composed of solitons could be detected by measures of substructure (which we discuss further below). DM composed entirely of solitons/oscillons/boson stars may have a number of interesting features for gravitational wave observations, as discussed in, e.g. \cite{2012arXiv1202.0560E,Macedo:2013jja}, that distinguish soliton DM from an equivalent model with black holes, due to the absence of a horizon. The soliton content of an axion DM halo will also impact the direct detection prospects \citep{Hoskins:2011iv}, while the evaporation of solitons from axion self interactions enhanced by the high number densities could have indirect signatures. Other signatures of the soliton components of an axion halo, e.g. in precision time-delay experiments with atomic clocks, may be similar to those with topological defect DM \citep{2013PhRvL.110b1803P,2014NatPh..10..933D,2014PhRvL.113o1301S}.

There are also `intermediate' cases to consider. ULAs with $m_a\lesssim 10^{-23}\text{ eV}$ could be detected as a sub-dominant component of the DM with high-precision studies \citep{Hlozek:2014lca}. A sub-dominant component at these masses is likely the expectation based on string models with sub-Planckian axion decay constants \citep[although see][]{Bachlechner:2014gfa}. ULAs with $m_a> 10^{-22}\text{ eV}$ as the dominant DM could also play some role in core formation alongside inefficient baryonic processes.

Finally, we turn to discussion of the optimistic case, where axions at the edge of our allowed region in mass are solely responsible for dSph cores. In this case, soliton cores have a large mass $\mathcal{O}(10^8\, M_\odot)$ and large radial extent $\mathcal{O}(1\text{ kpc})$, and the cut off in the halo mass function (in the field) occurs at $M\sim 10^8\, M_\odot$ \citep[the $z=0$ Jeans mass,][]{marsh2013b,bozek2014}. 

We have demonstrated the ability of soliton cores to fit the dSph observations, and suggested a complete density profile including the NFW piece. This density profile could be fit to many other observations, including rotation curves \citep{2001ApJ...552L..23D,2012MNRAS.422..282R} and stellar clump survival \citep{2014JCAP...09..011L}. We argue that further study of this model should use the density profiles we advocate in this work, and not `multi-state' or other adaptations of the scalar field DM model. The outer NFW piece obviates the need for such additions, and is consistent with many lines of reasoning. Further study should, as we have in this work, account for consistency of the cosmology, as well as between different density profile constraints. Consistency with cosmology and our core size fits in this work suggest that density profiles just outside 1 kpc should be much steeper than cores and make a transition to NFW shortly thereafter. It should also be investigated whether the scale symmetry for soliton cores can be used to explain the observed scaling relationships in dwarf galaxies \citep{2014arXiv1411.2170K,2015arXiv150106604B}.

While density profiles provide the motivation for $m_a\sim 10^{-22}\text{ eV}$, cosmology suggests that the most striking evidence for such a model will be found in sub-structure and high-$z$ galaxy formation. Measures of substructure from mili-lensing \citep{2002ApJ...572...25D}, tidal tails \citep{2002ApJ...570..656J}, and strong lensing \citep{2014arXiv1403.2720H} should be sensitive to the absence of small-scale structure predicted by an axion solution to the cusp-core problem. Further study is required to formulate the predictions of the axion model in this regard. This should come from simulation, but also from semi-analytic study of the conditional mass function and extended Press-Schecter formalism \citep{1993MNRAS.262..627L}. This study will further address the questions of whether axion DM can provide a consistent solution to other small-scale crises like `missing satellites.'

A final set of `smoking gun' signatures in the optimistic scenario are suggested by the results of \cite{bozek2014}. Since $m_a\sim 10^{-22}\text{ eV}$ is just consistent with the HUDF UV luminosity function, and also just able to provide a dSph core, this makes the prediction that a JWST measurement of the UV luminosity function sensitive to fainter magnitudes will only see a value consistent with HUDF, and not the larger value predicted by CDM. The cut off in the mass function also makes predictions about reionisation that are less sensitive to astrophysical uncertainties than in CDM. An axion solution to the cusp-core problem predicts that CMB polarisation experiments \citep{Calabrese:2014gwa} should measure: a low value of $\tau\sim 0.05$,\footnote{In \cite{bozek2014} it was noted that this is disfavoured at $\sim 2-3\sigma$ by WMAP \citep{2013ApJS..208...20B}, however it is in less tension with recent \emph{Planck} results \citep{planck_2015_params}.} a low redshift of reionisation $z_{\rm re}\sim 7$, and thin width of reionisation $\delta z_{\rm re}\sim 1.5$. These effects on reionisation, caused by the rapid build up of massive structures at the mass function cut-off, should also turn up striking effects in the 21cm power spectrum \citep[building on][]{2014JCAP...06..011K,2014MNRAS.438.2664S}. Further study of all of these signatures is underway.

\section*{Acknowledgments}

We thank Niayesh Afshordi, Mustafa Amin, Asimina Arvanitaki, Avery Broderick, Tzihong Chiueh, Malcolm Faribairn, Ren\'{e}e Hlozek, Andrew Liddle, Eugene Lim, Ue-Li Pen, Gray Rybka, David Spergel, Clifford Taubes, Luis Ure\~{n}a-Lopez and Rosemary Wyse for helpful discussions.  We especially thank Jorge Pe\~{n}arrubia for help using the data from \cite{Walker&Penarrubia2011}, the Royal Observatory of Edinburgh for hospitality, and Dan Grin for a careful reading of the manuscript. Research at Perimeter Institute is supported by the Government of Canada through Industry Canada and by the Province of Ontario through the Ministry of Research and Innovation.

\appendix

\section{The Schr\"{o}dinger-Poisson System}

\subsection{Derivation}
\label{appendix:derivation}

We begin by deriving our workhorse system, the Sch\"{o}dinger-Poission (SP) form of the field equations, following \cite{Widrow&Kaiser1993}. Further reading on this formalism can be found in \cite{Coles2002,Coles&Spencer2003,Johnstonetal2009}, and references therein.\footnote{Here, we are using a Schr\"odinger system as a change of variables to understand a classical wave equation, which also has a fluid description. Similarly, an understanding of classical fluids can help in understanding aspects of quantum mechanics. See \cite{park_quantum} for further discussion.} The SP system can be viewed as a fundamental picture of structure formation for axions or other scalar DM, or as a numerical procedure to model and study CDM with a suitable cut-off, having certain advantages over N-body simulations \citep{Uhlemann:2014npa}.

The Klein-Gordon (KG) equation for a free homogeneous scalar field $\phi$ of mass $m_a$, in an expanding universe with Hubble rate $H$, is:
\begin{equation}
\ddot{\phi} + 3H \dot{\phi} + \frac{c^4m_a^2}{\hbar^2} \phi = 0 \, . \label{eqn:EoM_units}
\end{equation}
From this point onwards, we will set $c=1$, but keep factors of $\hbar$ for now. In the regime $H \ll m$ (valid at late times, when the axion field is oscillating and behaving as DM) the equation above can be solved by WKB methods:
\begin{equation}
\phi = \frac{\hbar}{\sqrt{2} \, m_a} (\psi \, e^{-imt/\hbar} + \psi^* e^{imt/\hbar}) \, , 
\label{eqn:Widrow_ansatz}
\end{equation}
for the function $\psi$ that is slowly varying with time, i.e. $|m \, \psi| \gg |\hbar \, \dot{\psi}|$.

The time-time metric component is (in the Newtonian gauge and in physical time):
\begin{equation}
g_{00} = - [1+2V(r)] \, ,
\end{equation}
where $V(r)$ is the Newtonian potential. Treating this perturbatively, the KG equation for the inhomogeneous field becomes:
\begin{equation}
\frac{1}{\sqrt{-g}} \partial_\mu \left[ \sqrt{-g} g^{\mu\nu} \partial_\nu \right] \phi - \frac{m_a^2}{\hbar^2} \phi = 0  \, ,
\end{equation}
\begin{equation}
\ddot{\phi} + 3H\dot{\phi} - \partial^i \partial_i \phi + (1+2V) \frac{m_a^2}{\hbar^2} \phi =0 \, .
 \label{eqn:Klein-Gordon}
\end{equation}
In the non-relativistic limit, this equation has the same ansatz solution as the homogeneous field equation. Plugging in our ansatz and neglecting terms of order $O(\ddot{\psi})$, since $\psi$ is a slowly varying function of time, we find 
\begin{align}
\ddot{\phi} = &\sqrt{2}i \left( -\dot{\psi} e^{-im_at/\hbar} + \dot{\psi}^* e^{im_at/\hbar} \right) \nonumber \\
&-  \frac{m}{\sqrt{2}\hbar} \left( \psi \, e^{-im_at/\hbar} + \psi^* e^{im_at/\hbar} \right) \, ,
\end{align}
\begin{equation}
\nabla^2 \phi = \frac{\hbar}{\sqrt{2}m_a} \left( \nabla^2 \psi \, e^{-im_at/\hbar} + \nabla^2 \psi^* e^{im_at/\hbar} \right) \, .
\end{equation}
In the regime $H \ll m$, the second term in Eq.~(\ref{eqn:Klein-Gordon}) can be neglected. This gives:
\begin{align}
&\sqrt{2} \, i\, (- \dot{\psi} \, e^{-im_at/\hbar}  \!+\! \dot{\psi}^* e^{im_at/\hbar}) -\nonumber\\
&\frac{\hbar }{\sqrt{2} m} \left( \nabla^2 \psi \, e^{-im_at/\hbar} \!+\! \nabla^2 \psi^* e^{im_at/\hbar} \right)+ \nonumber\\
& 2V \frac{m}{\sqrt{2}\hbar} \left( \psi \, e^{-im_at/\hbar} + \psi^* e^{im_at/\hbar} \right) = 0 \, .
\end{align}
Multiplying the equation above by $\hbar/\sqrt{2}$ and identifying all terms with the same exponential factor, we find the familiar Schr\"odinger equation:
\begin{equation}
i\hbar \dot{\psi}  = - \frac{\hbar^2}{2m}\nabla^2 \psi + m_aV \psi \, . \label{eqn:above1}
\end{equation}
In the sub-horizon, non-relativistic limit, in any gauge, the Poisson equation is:
\begin{equation}
\nabla^2 V = 4 \pi G \rho = 4 \pi G \psi \psi^* \, . \label{eqn:above2}
\end{equation}
Eqs. (\ref{eqn:above1}) and (\ref{eqn:above2}) form the coupled SP system for inhomogeneous scalar fields, give as Eq.~(\ref{eqn:SP_system}). The non-relativistic limit in this case caused us to drop the gradient energy from the right hand side of the Poisson equation, setting $\rho=|\psi|^2$. This is valid in the limit that $k/m_a\ll 1$ for wavenumber $k$, i.e. $\partial_x \phi/(M_{pl}m_a)\ll 1$. Consistent solutions to the SP system must respect this limit.

In deriving this form of the field equations, the only quantity that has been treated perturbatively is the potential, $V$, i.e. we are working in the weak-field, Newtonian limit of General Relativity. We have not treated the field, $\phi$, or the density, $\rho$, perturbatively, and therefore our results will be valid in the non-linear (in terms of density fluctuations about the critical density) regime of gravitational collapse. The only clear limitation of our solutions will be instability to black hole formation: the Jeans limit \citep[for boson stars, this is the analogue of the Chandrasekar limit, see][for more details]{1969PhRv..187.1767R,1985MNRAS.215..575K,Seidel:1990jh,Choptuik:1992jv}.

The SP system is simply another way of rewriting the KG equation, and is related to the more familiar fluid treatment of scalar fields in cosmology \citep[e.g.][]{hu1998b}. The standard fluid picture is most useful for studying linear perturbations in Fourier space via the sound speed and Jeans instability \citep[e.g.][]{Hwang&Noh2009,2013PhLB..726..559N,Hlozek:2014lca,2015arXiv150106918A}. We will find the Schr\"{o}dinger picture to be more intuitive and useful for applications in real space relevant to non-linear densities in halos. 

\subsection{Dimensionless Variables}
\label{appendix:dimensions}

Plugging the soliton ansatz, Eq.~(\ref{eqn:soliton_ansatz}), into the SP system, Eqs.~ (\ref{eqn:SP_system}), yields:
\begin{align}
\nabla^2 \chi &= \frac{2m_a^2}{\hbar^2}(V - \hbar \frac{\gamma}{m_a}) \chi  \, , \nonumber \\
\nabla^2 V &= 4 \pi G \chi^2 \, .
\end{align}
From Eq.~(\ref{eqn:Widrow_ansatz}), we know the dimension of $[\chi] = [\psi] = [m_a][\phi] = [m_a]^2$, and we will now replace $\chi$, $r$ and $\gamma$ by the dimensionless variables $\tilde{\chi}$, $\tilde{r}$ and $\tilde{\gamma}$:
\begin{align}
\chi &\rightarrow \tilde{\chi} \, m_a M_{Pl} \sqrt{2}\left(\frac{c}{\hbar}\right)^{3/2}\rightarrow \tilde{\chi} m_a \frac{c^2}{\hbar}\frac{1}{\sqrt{4\pi G}} \, , \nonumber\\
r &\rightarrow  \tilde{r} \, \hbar /m_ac \, , \nonumber \\
\gamma &\rightarrow  \tilde{\gamma} \, m_a/\hbar \, .
\end{align}
In these dimensionless variables the SP system is:
\begin{align}
\tilde{\nabla}^2 \tilde{\chi} &= 2\left(V - \tilde{\gamma}\right) \tilde{\chi}\, , \\
\tilde{\nabla}^2V &= \tilde{\chi}^2 \, .
\end{align} 
Tildes can then be dropped.

\subsection{Scaling Relations}
\label{appendix:scaling}

The SP system of Eqs.~\ref{eqn:SP_System2} obeys a scaling relation $(r, \chi, V,\gamma) \rightarrow
(r/\lambda, \lambda^2 \chi, \lambda^2 V,\lambda^2\gamma)$ for scale factor $\lambda$ \citep{1969PhRv..187.1767R}. This is easy to check:
\begin{align}
\nabla^2 \chi &= \frac{1}{\lambda^2} \nabla^2 \tilde{\chi} = \frac{1}{\lambda^4} \tilde{\nabla}^2 \tilde{\chi} \, , \nonumber\\
2(V-\gamma)\chi &= 2\frac{(\tilde{V} - \tilde{\gamma})}{\lambda^2} \frac{\tilde{\chi}}{\lambda^2} \, , \nonumber\\
\nabla^2\chi &= 2(V - \gamma)\chi \, , \nonumber\\
\Leftrightarrow \; \; \frac{1}{\lambda^4} \tilde{\nabla}^2\tilde{\chi} &= \frac{2}{\lambda^4}(\tilde{V}-\tilde{\gamma})\tilde{\chi}\, , \nonumber\\
\Leftrightarrow \; \;  \tilde{\nabla}^2\tilde{\chi} &= 2(\tilde{V}-\tilde{\gamma})\tilde{\chi} \, . \nonumber
\end{align}
We can also extend this scaling relation to include the mass of the soliton:
\begin{align}
M_s &= \int_0^{r_s}4\pi r^2 \rho \, dr = \int_0^{r_s}4\pi r^2 |\chi|^2 dr  \nonumber\\
 &= \int_0^{\tilde{r}_s}4\pi r^2 \frac{|\tilde{\chi}|^2}{\lambda^4} \lambda^3 d\tilde{r} = \frac{\tilde{M_s}}{\lambda} \, ,\label{eqn:MassScaling}
\end{align}
and the density of the soliton:
\begin{equation}
\rho_{\rm sol} = |\chi|^2  = \frac{1}{\lambda^4}|\tilde{\chi}|^2 = \frac{\tilde{\rho}_{\rm sol}}{\lambda^4}
\end{equation}
Therefore, we can summarize the scaling relation as: 
\begin{equation}
(r, \chi, V, \gamma, M_s, \rho_s) \rightarrow
(r/\lambda, \lambda^2 \chi, \lambda^2 V, \lambda^2\gamma,\lambda M_s, \lambda^4 \rho_s) \label{eqn:scalingRelation}
\end{equation}

This scaling symmetry is very powerful, and makes results about the small-scale clustering of axion/scalar-field DM have a universal character. The scaling can be used to find the density profiles on different length and density scales at fixed axion mass, $m_a$, but can also be used to rescale the mass at fixed length scale. The scaling symmetry will be useful to us in Appendix~\ref{sec:schrodinger_numerical}, where we numerically find soliton solutions with arbitrary central density $\rho$ and then scale them to astrophysical densities.

For the SP system to be used as a model for the gravitational collapse of non-relativistic DM axions, its solutions must be consistent limits of the full KG equation satisfying $\gamma\ll 1$ and $k/m\ll 1$. Solutions to the SP system for a fixed scaling are formally solutions regardless of the value of $\gamma$ and need not satisfy these constraints. However, when we fix the scaling for real astrophysical systems, the rescaled values must satisfy these constraint. In typical DM halos, the virial velocity is $v\sim 100\text{ km s}^{-1}\ll c$. Since both $\gamma$ and $k$ are related to the kinetic energy, we therefore expect to find these limits satisfied in astrophysical objects. In our explicit examples of density profiles for dSphs this is indeed the case.

\subsection{Power Law Solutions}
\label{sec:cores_indices}

In this subsection, we are especially interested in the oscillaton profile at small radii. In the limit $ r \rightarrow 0$, we can assume that the dominant term in the series expansion of the solution $\chi(r)$ is of the form $\chi = C r^{\beta}$, for a given constant $C$ and exponent $\beta$ to be determined below. 

In the case $\beta \neq \{-1,0\}$ the Schr\"odinger equation implies:
\begin{align}
\chi'' + 2\frac{\chi'}{r} &=  \beta(\beta + 1) C r^{\beta-2} \, , \nonumber \\
&= 2(V-\gamma) C r^\beta \, .
\end{align}
Therefore, as $r \rightarrow 0$, we find that $V(r)$ obeys the equation:
\begin{equation}
V - \gamma = \frac{\beta(\beta+1)}{2r^2}
\end{equation}

Substituting into the Poisson equation:
\begin{align}
V'' + 2\frac{V'}{r} &= C^2 r^{2\beta} \nonumber \\
3\beta \frac{(\beta+1)}{r^4} - 2\beta \frac{(\beta+1)}{r^4} &= C^2 r^{2\beta} \nonumber \\
\beta(\beta+1) &= C^2 r^{2\beta+4} \nonumber 
\end{align}
we find that the SP system has a solution for $\beta = -2$. This implies a steeply increasing central density profile:
\begin{equation}
\rho(r) = \chi^2(r) \sim r^{-4}
\end{equation}
However, in this case, we can see that the mass within any radius $r_c$ diverges:
\begin{equation}
M = \int_0^{r_c} \rho(r) 4 \pi r^2 dr \sim 4 \pi \left( - \frac{1}{r} \right)\bigg|_0^{r_c} \rightarrow \infty
\end{equation}
Therefore, the requirement that the central mass is finite excludes the case $\beta = -2 \,$.

Next we check the cases $\beta = 0$ and $\beta = -1$ separately. 

For $\beta = -1$, we have that $\chi = C/r$, $\chi' = -C / r^2$, $\chi'' = 2C/r^3$, and thus:
\begin{equation}
\chi'' + 2\frac{\chi'}{r} = 2\frac{C}{r^3} - 2\frac{C}{r^3} = 0 = 2(V-\gamma) \frac{C}{r} \, . \nonumber
\end{equation}
This expression must be true as $r \rightarrow 0$, and therefore, it imposes that $\lim_{r\rightarrow0} V=\gamma$.
In this case, the Poisson equation becomes: 
\begin{equation}
 V'' + 2\frac{V'}{r} = \frac{C^2}{r^2} \, , \nonumber
 \end{equation} 
with $V = log (r)$ a solution of the equation, for a constant $C = \sqrt{3}$. 

Thus, the SP system allows solutions with $\beta = -1$, i.e. solutions of the form $\lim_{r \rightarrow 0} \chi = \sqrt{3}/r$. Unlike the $\beta = -2$ case, these solutions will have a finite mass. Nevertheless, solutions with $\beta = -1$ are non-physical because they correspond to infinite total energy:
\begin{align}
E =& \int_0^{r_c} (\nabla \chi)^2 4 \pi r^2 dr \nonumber \\
 =& \int_0^{r_c} \frac{3}{r^4} 4 \pi r^2 dr \sim 12 \pi \left(-\frac{1}{r}\right) \bigg|_0^{r_c} \rightarrow \infty \, . \nonumber
\end{align}

Finally, for $\beta = 0$, we must also consider the next terms in the expansion of $\chi (r)$ at small radii (i.e. $\chi(r) = C_0 + C_1 r^{\beta_1} + ... $ with $\beta_i >0$). If all the coefficients of terms with positive powers of $r$ are zero, we recover the trivial solution $\chi(r) = C$ at all radii. This trivial homogeneous solution corresponds to the background density and the global ground state. We can also have non-trivial solutions for which the dominant term in the expansion at $r \rightarrow 0$ is a constant and the higher order terms have non-zero coefficients. The numerical solution to the boundary value problem with $V(r\rightarrow \infty)=\chi(r\rightarrow \infty)=0$ found in Appendix~\ref{sec:schrodinger_numerical} is such a solution. It has $\lim_{r\rightarrow 0} \chi' = 0 \,$, while the boundary conditions forbid the global ground state and enforce $\lim_{r\rightarrow 0} \chi'' < 0 \,$.

The solutions with $\beta=-1$ and $\beta=-2$ are clearly not solutions that minimise the total energy, so they cannot represent the local ground state. Also, each violates the non-relativistic condition, $k/m\ll 1$, which in dimensionless variables and spherical co-ordinates reads $\partial_r \chi\ll 1$. These solutions are not consistent limits of the fundamental Einstein-Klein-Gordon equations. The existence of these solutions does, however, point to the instability of the $r^0$ flat-core (pseudo) soliton. The instability, however, is a relativistic one, and the SP system is not suitable to analyse it. The solutions become relativistic before the $\beta=-1$ or $\beta=-2$ solutions are ever found, and a black hole will form for finite mass and energy. The flat-core soliton solution is unstable to black hole formation when the central field value exceeds $\phi(0)\gtrsim 0.3 M_{pl}$ \citep{Seidel:1990jh}. This fixes the maximum, Chandrasekar-like, soliton mass as a function of $m_a$ \citep{1969PhRv..187.1767R}.

The result that the consistent, non-relativistic density profiles are flat as $r\rightarrow 0$ is not surprising given that we were analysing solitons, which are by definition spatially confined, non-dispersive and non-singular solutions of a non-linear field theory. 

\subsection{Numerical Solutions}
\label{sec:schrodinger_numerical}
The Schr\"{o}dinger-Poisson (SP) system found in Eq.~(\ref{eqn:SP_System2}) does not have an analytic solution. We devote this section to finding the soliton profile for the SP system numerically.
\subsubsection{Boundary Conditions}
We begin by noting that the SP system consists of two second-order ODEs with an additional unknown parameter $\gamma$, and thus we need to set 5 boundary conditions such that the system is uniquely determined.
In Section~\ref{sec:cores_indices} for $\chi\sim r^\beta$ as $r\rightarrow 0$ we found that $\beta = 0$ and therefore, $\chi'(0) = 0$. In addition, we are always free to normalize $\chi$ such that $\chi(0)=1$, and then later restore units and use the scaling symmetry to ensure that $\chi(0)^2$ matches the desired central density. Two additional boundary conditions can be set by imposing a vanishing density and potential far away from the core (i.e. $\chi(r\rightarrow \infty) = 0$ and  $V(r\rightarrow \infty) = 0$). Thus, the first four boundary conditions arise naturally, and we are only left with the task of finding one more. This last condition arises from the fact that $\chi'(0) = \chi''(0) = 0$ when we have a flat core (i.e. $\beta = 0$). Then, by differentiating the Schr\"odinger equation  with respect to $r$, we have:
\begin{equation}
\chi''' + 2\frac{\chi''}{r} - 2\frac{\chi'}{r^2} = 2V' + 2(V-\gamma) \chi'
\end{equation}
\begin{equation}
\Rightarrow V'(r\rightarrow 0) = 0
\end{equation}

Thus, in order to find the soliton profile, we need to solve the boundary value problem:
\begin{align}
\chi'' + \frac{2\chi'}{r} &= 2(V-\gamma)\chi \\
V'' + \frac{2V'}{r} &= \chi^2 \label{eqn:BVP_system}
\end{align} 
with boundary conditions:
\begin{align}
\chi(0) = 1 \\
\chi'(0) = 0 \\
\chi(\infty) = 0 \\
V'(0) = 0  \\
V(\infty) = 0 \\
\end{align}

Boundary value problems (BVPs) which have one or more of the boundary conditions set at infinity are inherently difficult to solve numerically. One can attempt to replace the boundary at infinity by a boundary at $r=R \gg 1$, but this method will be very sensitive to the choice of $R$ and, in most cases, will fail to converge to the exact solution of the BVP. Instead, we note that $ \chi \, (r \rightarrow \infty) = 0$ and $V \, (r \rightarrow \infty) = 0$, and by assuming that $\gamma$ is not too small (i.e. $V(R) \ll |\gamma|$ for $R\gg 1$), we can approximate the Schr\"odinger-Poisson system as follows:
\begin{align}
\chi'' + 2\frac{\chi'}{r} &= 2(V-\gamma)\chi\\
\chi''(R \gg 1) &\approx  -2\, \gamma \, \chi \,(R\gg 1) 
\end{align}
so at large values of $R$:
\begin{equation}
\chi(R) \propto e^{-\sqrt{- 2 \gamma} R} \; \; \; \mathrm{for } \gamma < 0
\label{eqn:chi_gamma_approx}
\end{equation}
We expect $\gamma < 0$ because $\chi$ must decay at large distances, whilst for positive values of $\gamma$, $\chi$ would have an oscillatory behaviour. If we replace this expression for $\chi(R)$ in the Poisson equation, (\ref{eqn:BVP_system}), we find:
\begin{equation}
V''(R) + 2\frac{V'(R)}{R} \propto e^{-2\sqrt{-2 \gamma}R}
\end{equation}
introducing the notation $U \equiv V'$:
\begin{equation}
U'(R) + 2\frac{U(R)}{R} \propto e^{-2\sqrt{-2\gamma}R}
\end{equation}
\begin{align}
\frac{dU}{Ce^{-2\sqrt{2\gamma} R} - 2\frac{U}{R}} &= dR \\ 
-\frac{dU}{2U} &\simeq \frac{dR}{R}\text{, for}R\gg 1
\end{align}
This gives the behaviour of the potential at large distances:
\begin{equation}
U(R) = V'(R) \propto R^{-2}
\end{equation}
\begin{equation}
V(R) \propto -\frac{1}{R}
\label{eqn:v_large_r_approx}
\end{equation}
Using the approximations of Eqs.~(\ref{eqn:chi_gamma_approx}) and (\ref{eqn:v_large_r_approx}), we can now replace the two boundary conditions at infinity (i.e. $\chi(\infty) = 0$ and $V(\infty) = 0$) by:
\begin{align}
\chi'(R) &= -\sqrt{-2\gamma} \, \chi(R) \\
V'(R) &= -\frac{V(R)}{R}
\end{align}
for a large value of $R$. In practice, we will test different values of $R$ and find the smallest value for which the solution still converges.

\subsubsection{Solution of the Boundary Value Problem}
\label{subsec:SolitonNumAnalysis}

To solve the boundary value problem, convert the system of 2 second-order ODEs into a system of 4 first-order ODEs, by introducing the quantities $\mathcal{X} \equiv \chi'$ and $\mathcal{V} \equiv V'$:
\begin{equation}
\mathbf{y} =  \left( \begin{array}{c}
\chi\\
\mathcal{X}\\
V\\
\mathcal{V}\end{array} \right)
\end{equation}
\begin{equation}
\mathbf{y'} = \left( \begin{array}{c}
\chi'\\
\mathcal{X}'\\
V'\\
\mathcal{V}'
\end{array} \right) = \left( \begin{array}{c}
\mathcal{X} \\
-2\frac{\mathcal{X}}{r} + 2(V-\gamma) \chi \\ 
\mathcal{V} \\
-2\frac{\mathcal{V}}{r} + \chi^2
\end{array} \right)
\end{equation}
with the following five boundary conditions (noting again that the fifth boundary condition arises due to the presence of the unknown parameter $\gamma$):
\begin{align}
\chi(0) &= 1 \, , \nonumber \\
\chi'(0) &= 0 \, , \nonumber \\
V'(0) &= 0 \, , \nonumber \\
\chi'(R) &= -\sqrt{-2\gamma} \, \chi(R) \, , \nonumber \\
V'(R) &= -V(R)/R \, .
\end{align}

This system does not have a unique solution $(\mathbf{y},\gamma)$. Instead, it can be viewed as an eigenvalue problem with unique solutions $(\mathbf{y}_i, \gamma_i)$ for each eigenvalue $\lambda_i$ or number of nodes $n_i$. Over time, the system will relax into the stable state of minimum energy, which corresponds to a soliton with zero nodes (i.e. the ground state). Thus, we look not only for a solution of the BVP that converges, but also for the exact solution $(\mathbf{y}, \gamma)$ which has zero nodes.

\begin{figure}
\centerline{\includegraphics[width = .5\textwidth]{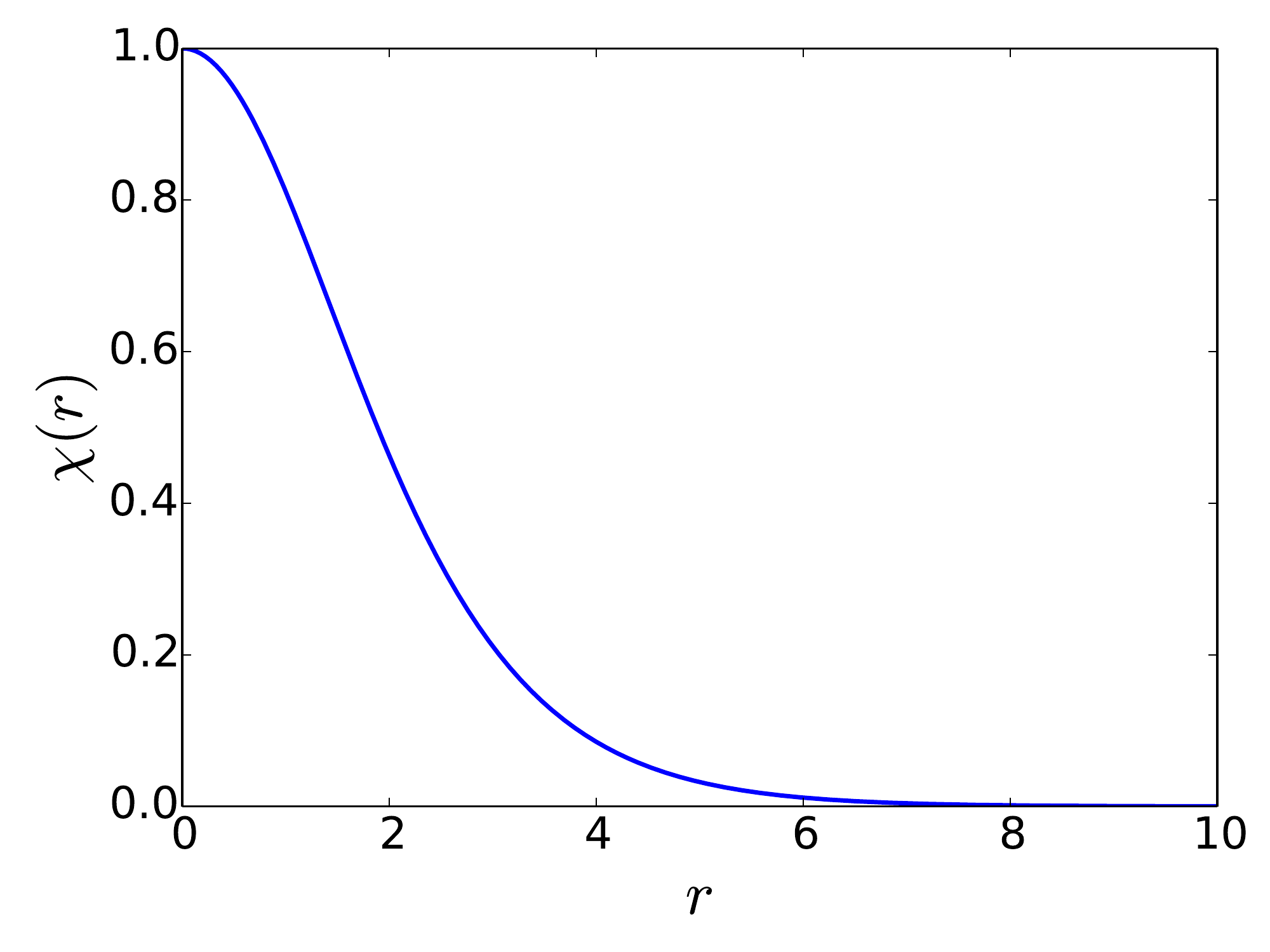}}
\caption{Zero-node soliton profile $\chi(r)$ found by solving the BVP system in Eq.~\ref{eqn:BVP_system}. }\label{fig:Phi_Soliton}
\end{figure}

The zero node soliton profile $\chi(r)$ is depicted in Fig. \ref{fig:Phi_Soliton}. We find that the zero-node solution presented in this figure corresponds to:
\begin{align}
\gamma &= -0.692 \, , \\
V(r=0) &= -1.341 \, .
\end{align}
These values are in agreement with those reported in Table II of \cite{Guzman&Lopez2004}, who studied an equivalent system. We note that our soliton solution has $|\gamma|<1$ and $|\partial_r \chi|<1\, \forall r$. They are consistent limits of the Einstein-Klein-Gordon equations. The scale parameter $\lambda<1$ that takes them to astrophysical systems after restoring units reduces these values further and improves the validity of the non-relativistic, Newtonian limit.

\begin{figure}
\centerline{\includegraphics[width = .5\textwidth]{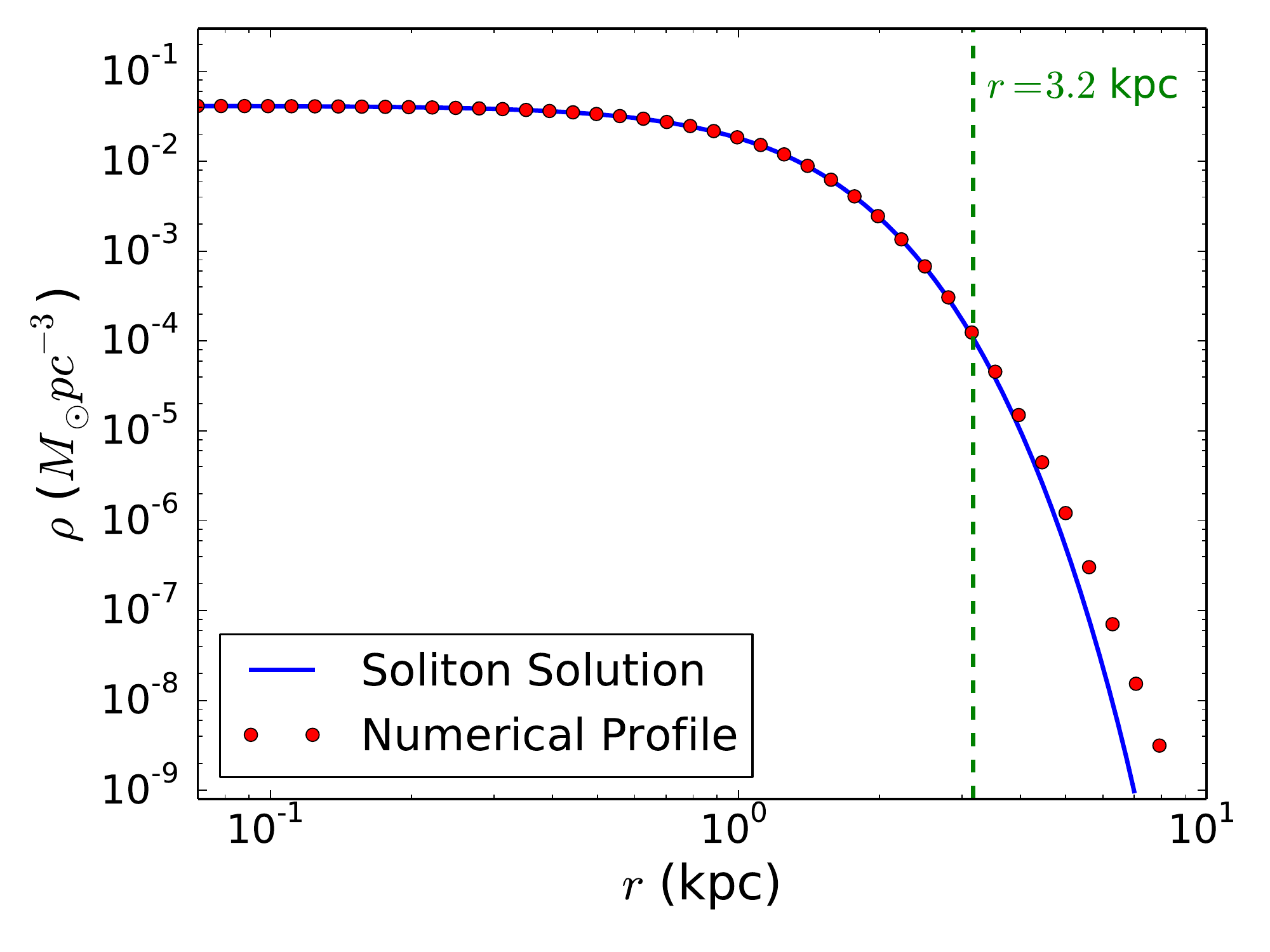}}
\caption{In blue is depicted the density profile obtained from the zero-node soliton solution, scaled to typical dwarf galaxy parameters, while in red we show the fit of Eq.~(\ref{eqn:Schive_rho}). This provides a good fit to the BVP solution up to radius $r \approx 3.2 \, \mathrm{kpc}$.}\label{fig:Phi_Soliton_convergence}
\end{figure}

\subsection{Fitting the soliton density profile} 
\label{appendix:fits}

We fit the ground state soliton density profile, $\tilde{\rho}=f(\alpha r)$ for different functional forms. We begin with the form used by \cite{2014NatPh..10..496S}: 
\begin{equation}
f(\alpha r) = \frac{1}{(1+\alpha^2 r^2)^8} \, ,
\label{eqn:polynomial}
\end{equation}
In \cite{2014NatPh..10..496S,Schiveetal2014b} the soliton-NFW transition generally occurs for $\rho\sim 0.01 \rho(0)$. Therefore we perform a best fit for the value of $\alpha$ with $r$ chosen so that $\tilde{\rho}\in[0.01,1]$. For this range of $r$ we find $\alpha=0.230$. This form has a number of advantages. It gives an analytic expression for the mass integrated out to any radius. Assuming this form for the density, one can also find an analytic solution for the Newtonian potential $V(r)$ from the Poission equation, and the combined solution can be shown to give an accurate solution to the full SP system. Allowing the exponents in the fit to vary can improve the fit slightly at large radius, but loses these computational advantages. In our complete profile the large radius region occurs after the transition to NFW and improving the soliton fit here does not have any effect.

We also investigate a Gaussian fit:
\begin{equation}
f(\alpha r)= \mathrm{exp}\left[-\frac{r^2}{2\sigma^2}\right] \label{eqn:Gaussian}
\end{equation}
and find $\sigma = \alpha^{-1}/\sqrt{2}= 1.22$. This form of the density profile also gives analytic results for the soliton mass, and of slightly simpler form than for the polynomial fit. However, we find that the Gaussian profile always gives a worse fit than the polynomial, and consider it no further. 

\cite{2014NatPh..10..496S} solved the system of ODEs in (\ref{eqn:SP_System2}) using the shooting method and the boundary conditions of \cite{Guzman&Lopez2006}:
\begin{align}
\phi(0) &= \partial_r \phi (0) = 0 \\
\phi (r \rightarrow \infty) &= 0 \\
V(\infty) &= 0
\end{align}
\cite{2014NatPh..10..496S}'s fit for the soliton density profile is:
\begin{equation}
\rho_{\rm sol} (r) \simeq \frac{1.9 \, (m_B/10^{-23} \mathrm{eV})^{-2} (r_{1/2}/\mathrm{kpc})^{-4}}{[1+9.1 \times 10^{-2} (r/r_{1/2})^2]^8} M_{\odot} \, \mathrm{pc}^{-3} \, , \label{eqn:Schive_rho}
\end{equation}
where they define $r_{1/2}$ such that $\rho_{\rm sol}(r_{1/2})=\rho_{\rm sol}(0)/2$. This fixes the factor $9.1\times 10^{-2}$ and the value of $\alpha$ is found by normalising the central density. Restoring units using the scaling symmetry, as outlined in Section~\ref{sec:uncertainty}, shows good agreement between this fit and ours. 

\section{A Note on Solitons}
\label{appendix:soliton_note}

We have considered spherically symmetric soliton solutions to Eqs.~(\ref{eqn:SP_system}). In this model, the non-linearity necessary to support solitons comes purely from gravity. Real fields such as axions form a class of solitons known as oscillons, which, being unprotected by a charge, are technically pseudo-solitons. Oscillatons are oscillons including self-gravity. The solutions we study evade Derrik's theorem that otherwise forbids solitons in three spatial dimensions because they are time-dependent and include self-gravity. Scalar field solitons known as boson stars are formed for complex fields, where stability is guaranteed by the charge. For more discussion of solitons, oscillons and boson stars in various contexts, see e.g. \cite{1969PhRv..187.1767R,Liddle:1993ha,Liebling:2012fv,2012PhRvL.108x1302A,Amin:2013ika,Lozanov:2014zfa}.

We considered the ground state solitons. An isolated system relaxes to the ground state via the emission of scalar waves to infinity, a process of `gravitational cooling' \citep{Seidel:1993zk,Guzman&Lopez2006}. Stable solutions are the non-linear density profiles that form the end-point of spherical collapse in axion DM halos below the threshold for black hole formation. These solutions are pressure supported. By analogy to stars, this justifies the use of spherical symmetry, as we expect the relaxation time for non-spherical perturbations to be on the same time-scale as the relaxation time to the ground state. The simulations of \cite{2014NatPh..10..496S} show that these ground state solitons form and are stable on cosmological time scales.  

During cosmological structure formation, such ground state solutions will only be found locally if the relaxation time is shorter than the age of the universe, with the relaxation time being shorter for denser objects. These solitons will possess a characteristic size related to their formation time. CDM-like structure will form on larger scales due to the Jeans instability \citep{1985MNRAS.215..575K}. Furthermore, in bound states such as halos, scalar waves may not escape to infinity and may contribute to the outer parts of a halo. Detailed questions about the non-linear dynamics can only be solved by simulation and will be the subject of future work.

\bsp

\label{lastpage}

\end{document}